\documentclass[twocolumn,showpacs,amsmath,amssymb,aps,prc]{revtex4-1}

\RequirePackage{lineno}
\usepackage{graphicx}
\usepackage{dcolumn}
\usepackage{bm}

\begin{document}

\preprint{APS/123-QED}

\title{Comment on ``Evidence for narrow resonant structures at {\boldmath$W\approx$}~1.68~GeV and
{\boldmath$W\approx$}~1.72~GeV in real Compton scattering off the proton''}

\author{
  D.~Werthm\"uller,$^{1,2}$
  L.~Witthauer,$^2$
  D.I.~Glazier,$^{1}$
  B.~Krusche$^2$
}

\affiliation{
  $^{1}$\mbox{School of Physics and Astronomy, University of Glasgow, Glasgow G12 8QQ, United Kingdom}\\
  $^{2}$\mbox{Departement Physik, Universit\"at Basel, CH-4056 Basel, Switzerland}\\
}

\date{\today}

\begin{abstract}
We comment on the statement by Kuznetsov {\it et al.} that the structure around $W=1.72$ GeV
seen in the beam asymmetry in Compton scattering off the proton is not observed in the
total cross section of $\eta$ photoproduction on the neutron.
\end{abstract}

\pacs{14.20.Gk, 13.60.Rj, 13.60.Le}

\maketitle

In a recent paper Kuznetsov et al.~\cite{Kuznetsov_15}, presented the results of Compton scattering off the proton
from the GRAAL experiment. The discussion focused on narrow peak structures in the beam asymmetry
$\Sigma$ for this reaction. They related these structures to earlier observations in photoproduction of 
$\eta$ mesons, in particular off neutrons, from the GRAAL \cite{Kuznetsov_07}, ELSA \cite{Jaegle_08,Jaegle_11}, 
and MAMI \cite{Werthmueller_13,Witthauer_13,Werthmueller_14} facilities.

\begin{figure}[b!]
\centering
\includegraphics[width=0.48\textwidth]{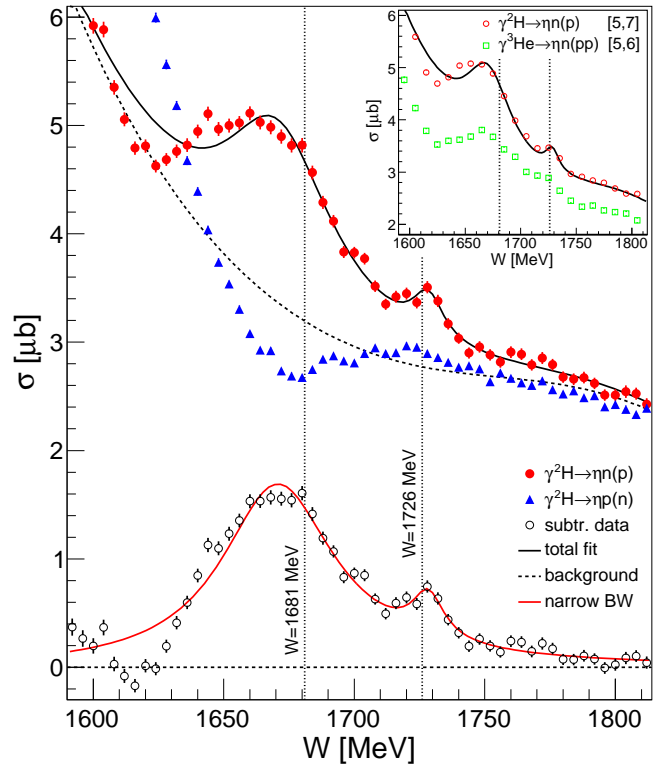}
\caption{(Color online) Excitation functions for quasifree photoproduction of $\eta$ mesons.
Inset: total cross sections for quasifree neutrons bound in the deuteron (red circles) and
in $^3$He nuclei (green squares). Data are as published in Fig.~1 of Ref.~\cite{Werthmueller_13}.
Fit curves include Breit-Wigner resonance for the structure at 1726~MeV.
Main figure: finer binned data for the measurement with deuterium target. Red circles (blue triangles): 
quasifree neutron (proton) data. Curves: total fit (solid black curve), background contributions 
(dashed black curve), and sum of two narrow signal Breit-Wigner functions (red solid curve).
Open black circles: background-subtracted neutron data.
Vertical lines: markers at $W=1681$ MeV and $W=1726$ MeV.}
\label{fig:result}
\end{figure}

We would like to comment on one important issue that has been incorrectly 
raised in their paper. The structures in Compton scattering have been observed at total 
center-of-mass (c.m.) energies of 1681~MeV and 1726~MeV. The peak at smaller $W$ has a
prominent counterpart in $\eta$ photoproduction off the neutron (although the actual 
peak positions from different measurement deviate by up to 15 MeV and the width is different
from the structure observed in Compton scattering off the proton). However, for the structure 
at 1726~MeV the authors note that it has not been observed in other reactions or observables,
and state in their conclusions: ``On the other hand it still remains unclear \dots\ why the structure at
$W\approx1.72$~GeV is seen in Compton scattering and is not seen in $\eta$ photoproduction on 
the neutron.'' This statement would be very important for the interpretation of the nature of the
structure, but it is not correct. Actually, at this c.m.\ energy there is a narrow,
small --- but statistically significant --- structure present in already published data 
\cite{Werthmueller_13,Witthauer_13,Werthmueller_14} for $\eta$ photoproduction off the neutron.
The little peak in the total cross sections in \cite{Werthmueller_13,Witthauer_13,Werthmueller_14}
was not explicitly discussed in these references (and is better visible in a finer binned
version of these data) so that it has obviously escaped attention of the authors
of \cite{Kuznetsov_15}.

In order to demonstrate this, we show in Fig.~\ref{fig:result} the excitation functions for 
photoproduction of $\eta$ mesons off neutrons bound in the deuteron and in $^3$He nuclei. 
The results published in \cite{Werthmueller_13} are shown in the small inset with the relevant
energy range magnified. Both excitation functions for the neutron show a statistically significant, 
although small peak at exactly the same invariant mass as the structure observed in the beam 
asymmetry of Compton scattering off the proton. The agreement between the measurements with the 
deuterium and the $^3$He targets (the signals agree in position and are of comparable size and width)
suggest that this signal is robust. The main part of the figure shows the excitation functions for 
quasifree neutrons and protons bound in the deuteron with a finer energy binning. The neutron data
were fitted with the same parametrization as used in \cite{Werthmueller_14} adding an additional 
Breit-Wigner function to account for the structure near $W=1720$~MeV. The position was determined to 
be $M = (1728\pm 1)$~MeV and a width of $\Gamma = (15\pm 4)$~MeV was extracted. The indicated 
uncertainties are only statistical.

The nature of these structures and the relation between the results from Compton scattering off 
the neutron \cite{Kuznetsov_11} and the proton \cite{Kuznetsov_15} and $\eta$ photoproduction off 
the neutron \cite{Kuznetsov_07,Jaegle_08,Jaegle_11,Werthmueller_13,Witthauer_13,Werthmueller_14} 
are not yet understood. For $\eta$ production, the prominent structure at $W\approx1670$~MeV
has been recently interpreted in the framework of the Bonn-Gatchina (BnGa) partial wave analysis 
mainly as an interference pattern in the $S_{11}$ partial wave \cite{Anisovich_15}.
The interference between the $N(1535)1/2^-$ and $N(1650)1/2^-$ states changes sign between protons and 
neutrons so that it is destructive for the proton and constructive for the neutron. The comparison 
of proton and neutron excitation functions in this energy range (see Fig.~\ref{fig:result}) is indeed 
suggestive for such an interference pattern. The structure observed in Compton scattering off the
proton \cite{Kuznetsov_15} is narrower and peaks at a larger invariant mass 
(position indicated in Fig.~\ref{fig:result}) which coincides with a little dip in the proton
cross section for $\eta$ photoproduction.

The peak observed around $W\approx 1720$~MeV in the beam asymmetry of the 
\mbox{$\gamma p\rightarrow p\gamma '$} reaction coincides
exactly with the clear, statistically significant structure observed in $\eta$ photoproduction 
with coincident detection of recoil neutrons from deuterium and $^3$He nuclei. The data with 
coincident recoil protons may show a tiny structure at this invariant mass, but this is at 
the very edge of statistical significance. The position of this structure is suspiciously close 
to the production threshold of $\omega$ mesons off the free neutron, $W_{\rm thr}\approx 1722$~MeV.
As already noted in Ref.~\cite{Kuznetsov_15}, a detailed combined analysis of $\eta$ photoproduction 
and Compton scattering off protons and neutrons in this energy range seems to be highly desirable 
to clarify the nature of this rich spectral features.

In conclusion, contrary to the statements in the paper by Kuznetsov et al.~\cite{Kuznetsov_15}, experimental data for 
quasifree photoproduction of $\eta$ mesons off neutrons show a significant narrow structure 
at precisely the invariant mass at which the second narrow peak in the beam asymmetry $\Sigma$ for
Compton scattering off the proton has been reported. This signal in the $\gamma n\rightarrow n\eta$
reaction is much fainter than the prominent structure at $W\approx 1670$~MeV and has therefore
only been observed in the high statistics measurements with deuterium and $^3$He targets at MAMI
\cite{Werthmueller_13,Witthauer_13,Werthmueller_14}, but not in the previous 
GRAAL \cite{Kuznetsov_07} and ELSA \cite{Jaegle_08,Jaegle_11} experiments. 
So far it has not been discussed in any detail in the literature.

This work was supported by Schweizerischer Nationalfonds (200020-156983, 132799, 121781, 117601, 
113511 and 158822).

\end{document}